\documentclass{article}
\usepackage{graphicx} 

\title{On the `mosaic' picture of liquids and glasses
\footnote{An overview based on work in collaboration with Ittai Fraenkel and Dov Levine. Dedicated to the memory of G\'erard Toulouse.}
}

\author{Jorge Kurchan \\ Laboratoire de Physique de l'\'Ecole Normale Sup\'erieure, \\
ENS, Universit\'e PSL, CNRS, Sorbonne Universit\'e, \\Universit\'e de Paris, F-75005 Paris, France }

\date{June 2024}

\begin{document}

\maketitle

\begin{abstract}
    Supercooled liquids are sometimes described as being composed of  a mosaic of  patches that may be listed in a `library',  each one having some form of non-periodic order. Looking closer, one finds this construction elusive. In attempting to give the notion of mosaic a precise sense, we find that we are inevitably led to  the construction of a procedure for compressing the information in the particle configuration, essentially the same as that used  for texts. The amount of optimally  stored information directly defines the configurational entropy. A solid, in this view, is a particle arrangement described by a low amount of information, that can only flow by breaking into uncorrelated pieces, thus increasing its complexity. 
\end{abstract}

\section{Introduction}

As we cool a liquid to lower and lower temperatures, the viscosity goes up dramatically, 
 and some two-time correlations develop a very slow relaxation.
Perhaps there is a divergence  at some temperature  -- this would be the `ideal glass transition': an interesting question that will not be our main concern here, and whose answer we are far from knowing.  `Ideal' then means that the timescales are ideally infinite, and the system is a true solid, rather than being one for all practical purposes.

\begin{figure}
    \begin{center}
        \includegraphics[angle=90,width=8cm]{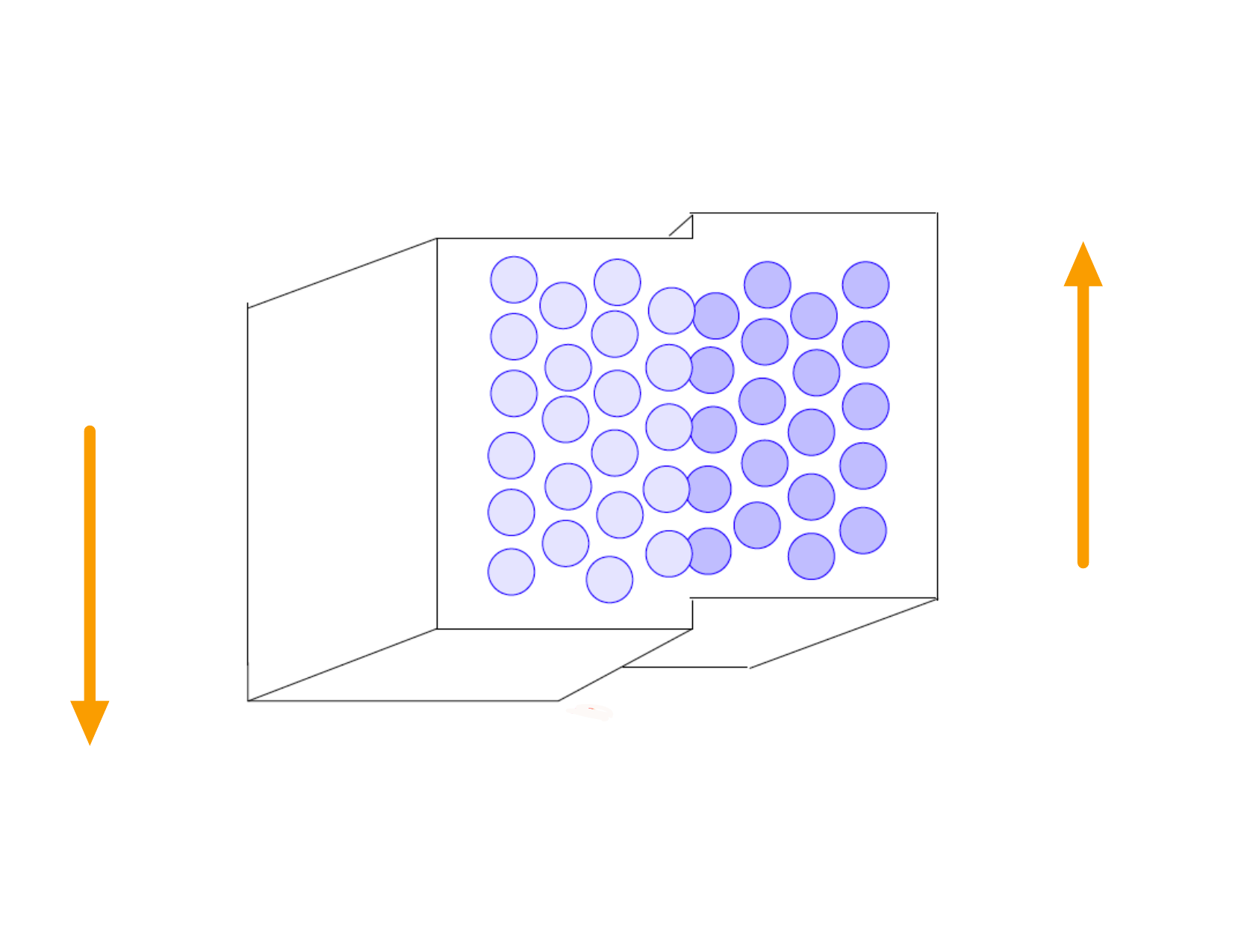}
   \caption{Thanks to its order, a sheared crystal made of soft particles develops an infinite energy barrier in the thermodynamic limit.}  \label{shear}\end{center}
\end{figure}

\begin{figure}
    \begin{center}
        \includegraphics[angle=90,width=8cm]{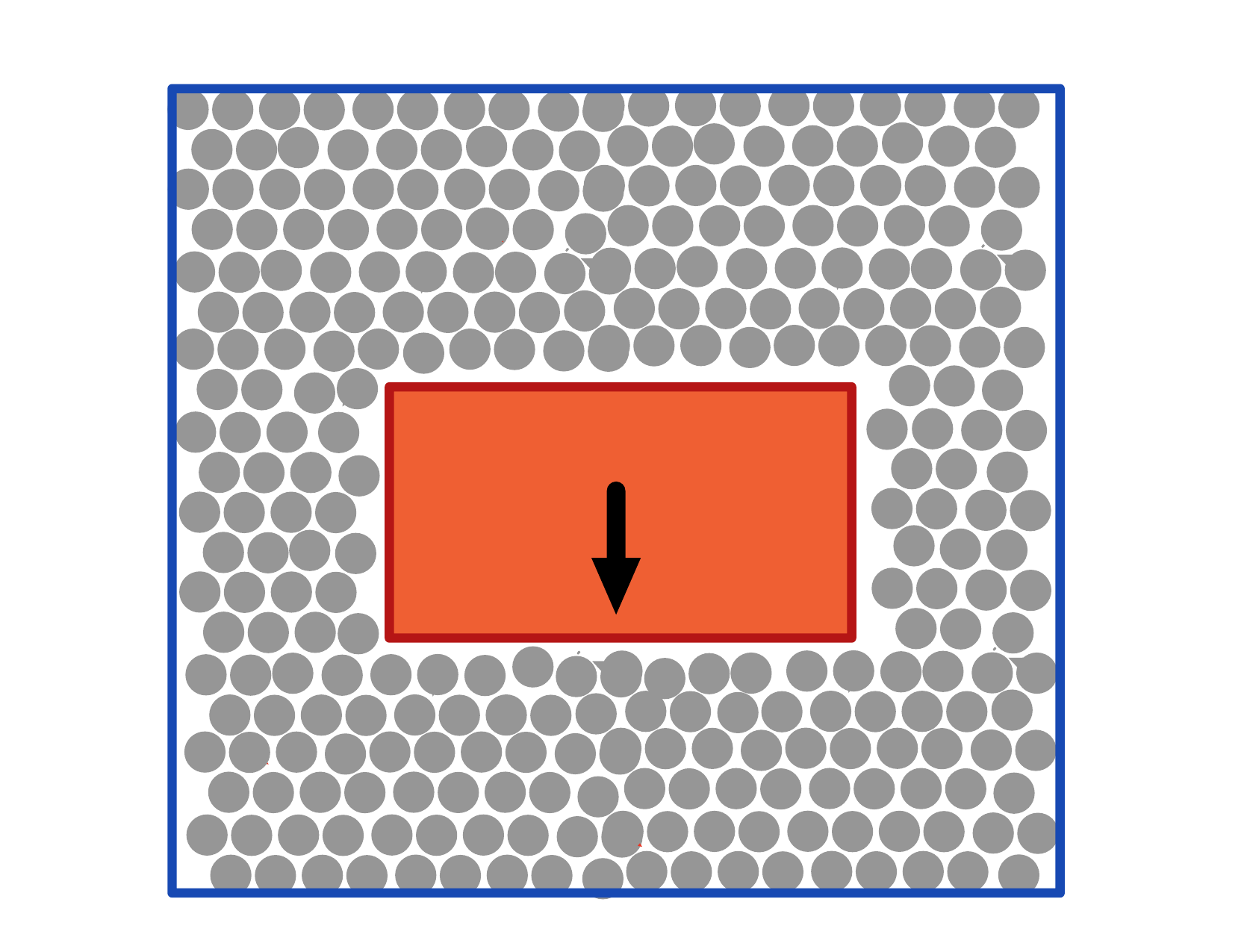}
    \caption{A solid sustains shear thanks to (some for of) order.} \label{budda} \end{center}
\end{figure}

The development of solidity, even an imperfect one as in a supercooled liquid, is a highly nontrivial phenomenon \cite{anderson1972more}. To convince oneself of this, it is better to consider particles that have no hard core and are thus interpenetrable. Now, these are known to form both crystalline solids and glasses, showing
that the image of a solid being made of particles `sitting upon each other' is misleading, and that something more subtle is at work. What is happening is an order that develops, involving more and more particles (see figs \ref{shear} and \ref{budda}).


\begin{figure}
    \begin{center}
        \includegraphics[angle=90,width=8cm]{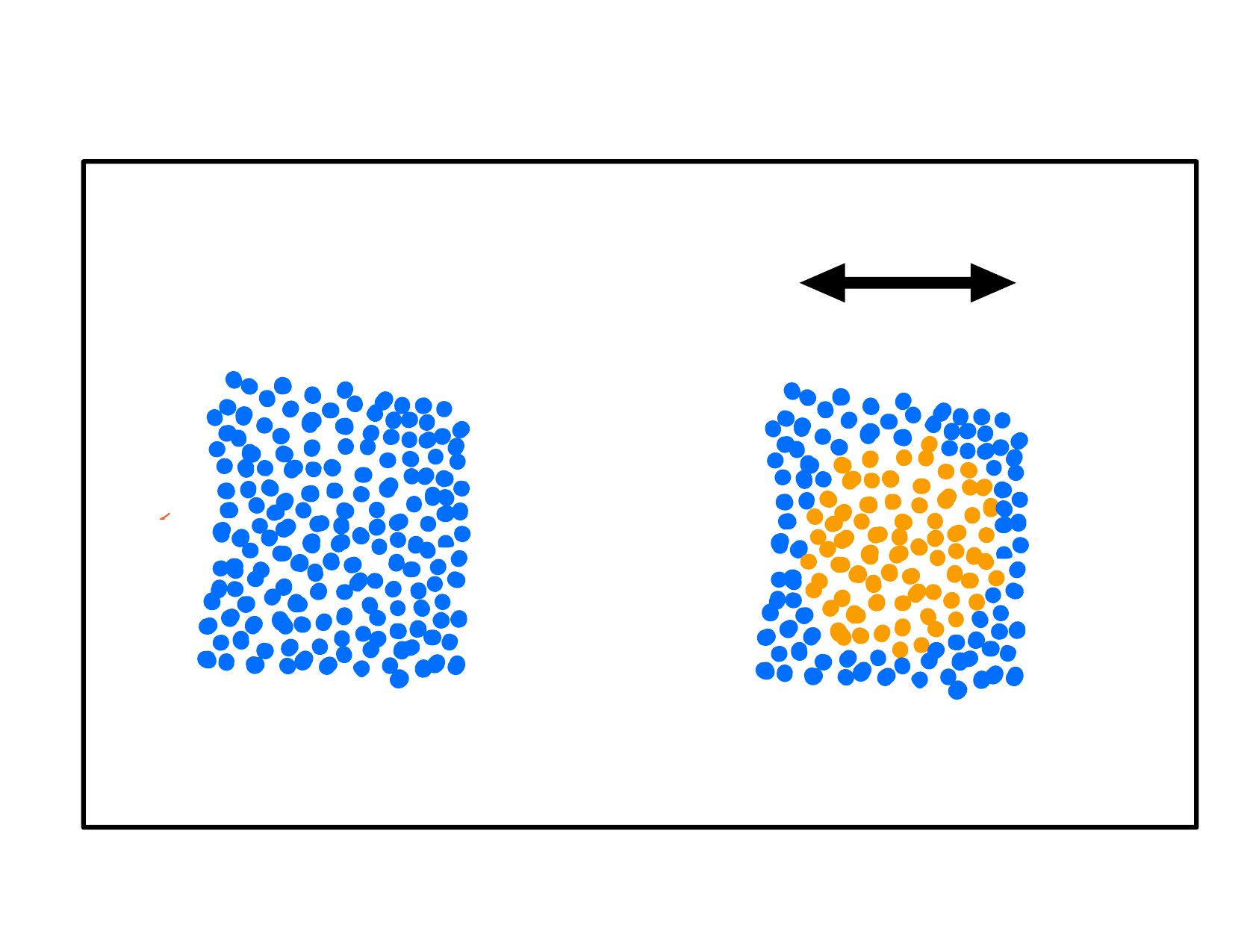}
    \caption{The point-to-set construction: the outside is frozen, the cavity is rethermalized} \label{pts}\end{center}
\end{figure}

A way to formalize this idea is to consider the   point-to-set construction\cite{biroli2004diverging,bouchaud2004adam,montanari2006rigorous}, as in Figure \ref{pts}.
We consider a supercooled liquid . At some time we freeze  (in the computer) all particles around a
sphere of radius $R$ and  we reshuffle the particles inside. If, upon letting them relax, they come
back to a configuration close to the original one, this tells us that the boundary determines the interior. We repeat for larger $R$ up to the point in which this no longer happens: the maximal length obtained this way is the `point-to-set length' $R_{ps}$.
 Now, it is clear that if the particles are soft, the barrier heights the system can construct at finite
 $R_{ps}$ are bounded by some function of $R_{ps}$ \cite{bouchaud2004adam,biroli2004diverging,montanari2006rigorous}. Our question is then to what extent does $R_{ps}$
 grow, and, a more academic one, if there is a true divergence.

 This kind of ideas has naturally led   to the picture in which a supercooled liquid
 - and its out of equilibrium version, a glass - are a mosaic \cite{lubchenko2007theory,xia2000fragilities,xia2001microscopic} of patches of ordered particles, `ordered' in the sense, for example, that they are the interior of a point-to-set cavity constructed above. These are not all equal up to rotations, as we would have in an assembly of microcrystals, and therein lies the problem. 

I will first describe how one makes a mosaic model meaningful. For simplicity, on a first approach (Sect \ref{mosaic}) I will neglect the important question of when do we consider two patches to be the same -- what  precision we require -- and how do we factor out the rapid thermal fluctuations (see Fig \ref{vibrations}). These issues will be dealt with in Section \ref{temp}.

\begin{figure}
    \begin{center}
        \includegraphics[angle=90,width=8cm]{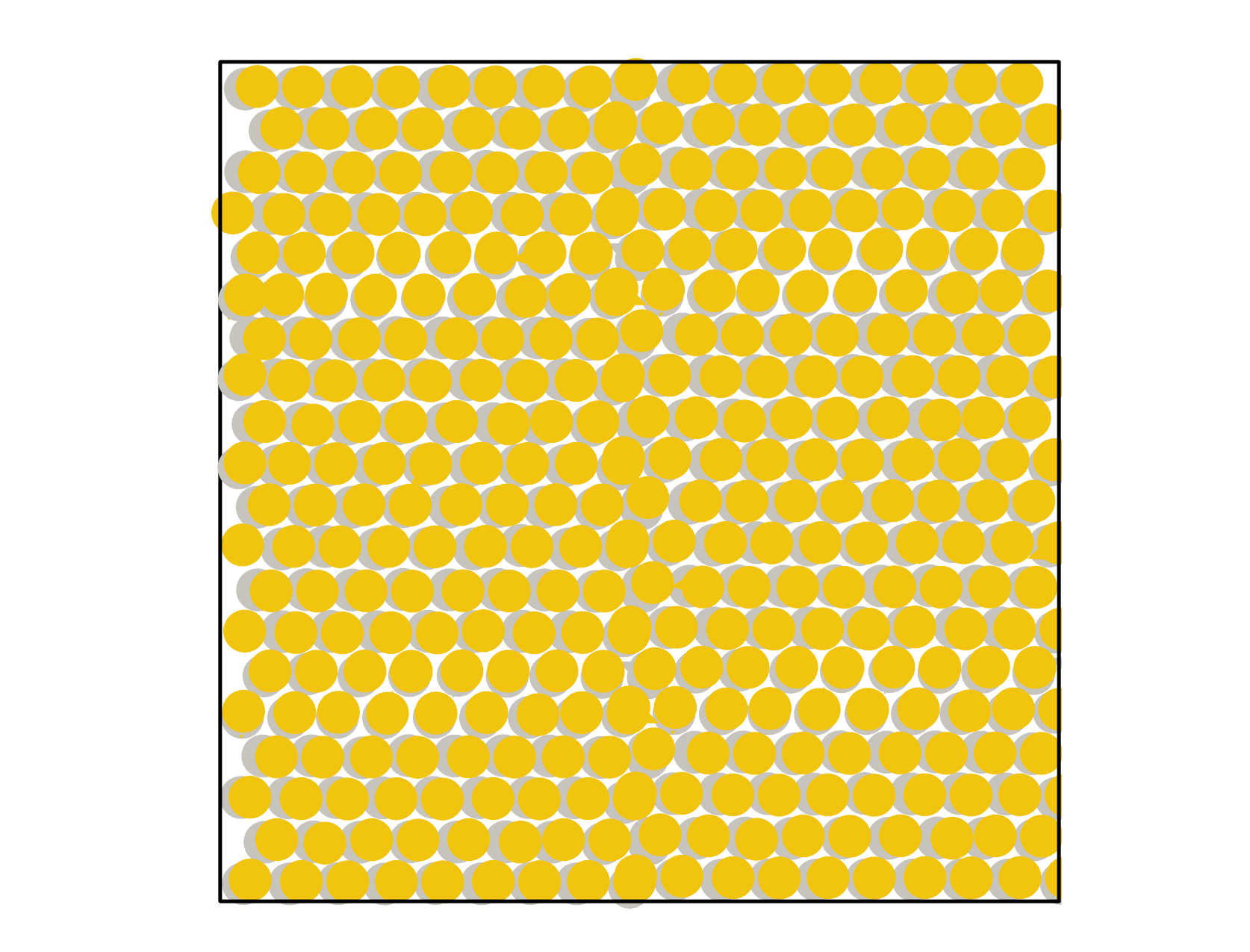}
    \caption{Vibration around an ordered system \label{vibrations}}  \label{vibrations}\end{center}
\end{figure}
The approach described here, and the inspiration coming from the Random First Order scenario \cite{kirkpatrick1987p,kirkpatrick1989scaling,wolynes1997entropy,xia2000fragilities,xia2001microscopic,lubchenko2007theory} takes us directly to a spatial characterization of effective temperatures \cite{cugliandolo1997energy} (Sect \ref{renyi}) and the Parisi-Monasson  $X$  parameter \cite{monasson1995structural}  in terms of the Renyi entropies that measure which ordered patches are typical and which are rare. 
It  turns out that a supercooled liquid has the whole repertoire of  ordered patches, some more rare and some more common, as temperature (or time, in an aging system) is changed, rare patches become typical and typical rare. 

This   overview is based on  work with Dov Levine and Ittai Fraenkel\cite{kurchan2009correlation,kurchan2010order,fraenkel2024information}. 

\section{Mosaics \label{mosaic}} 

\subsection{A misleading picture}

\begin{figure}
    \begin{center}
        \includegraphics[angle=90,width=8cm]{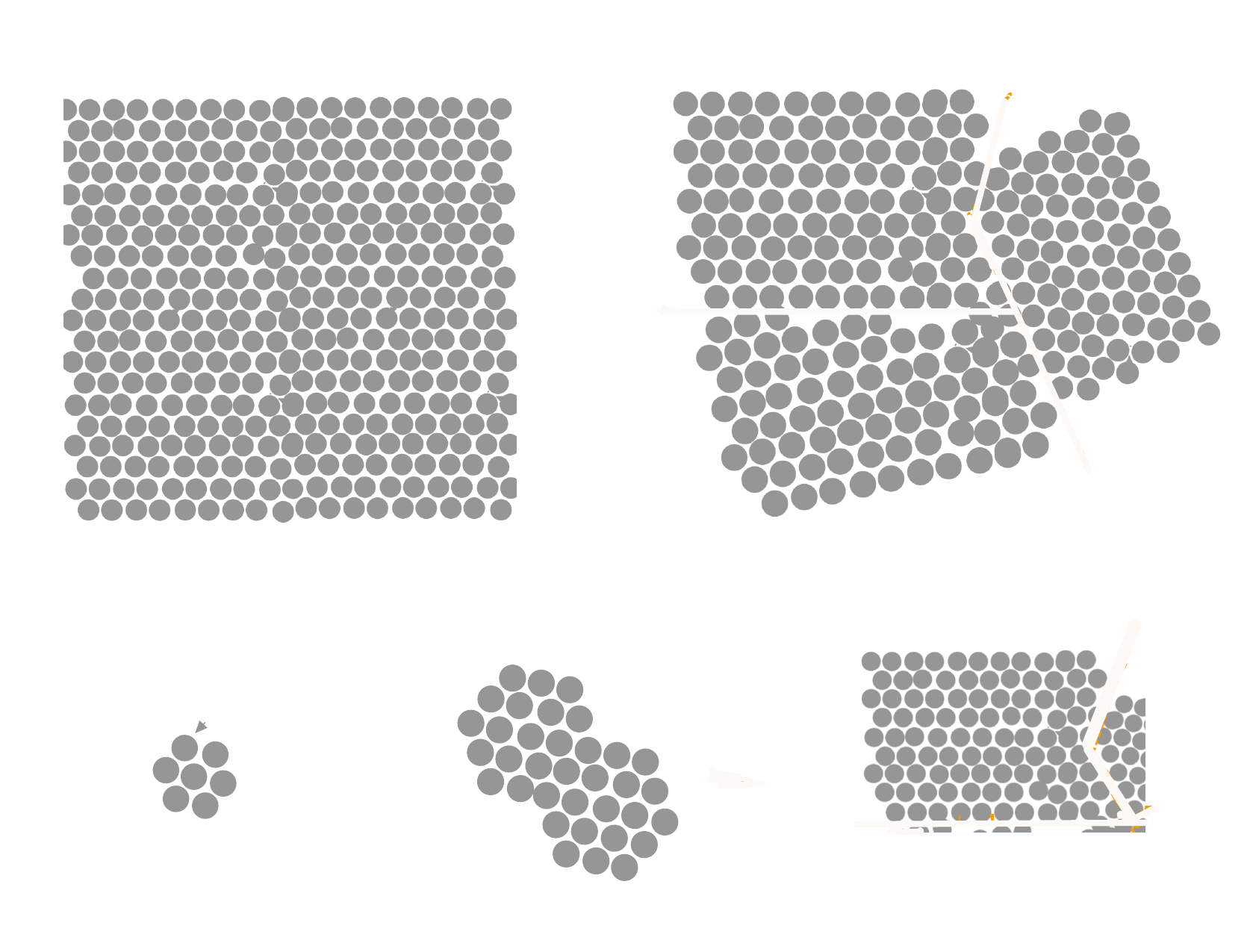}
    \caption{Top left: a crystal, right: three microcystallites. Below: a tile, two tiles make a new tile, two incoherent ones do not.  } \end{center} \label{microcrystal}
\end{figure}
\begin{figure}
    \begin{center}        \includegraphics[angle=90,width=8cm]{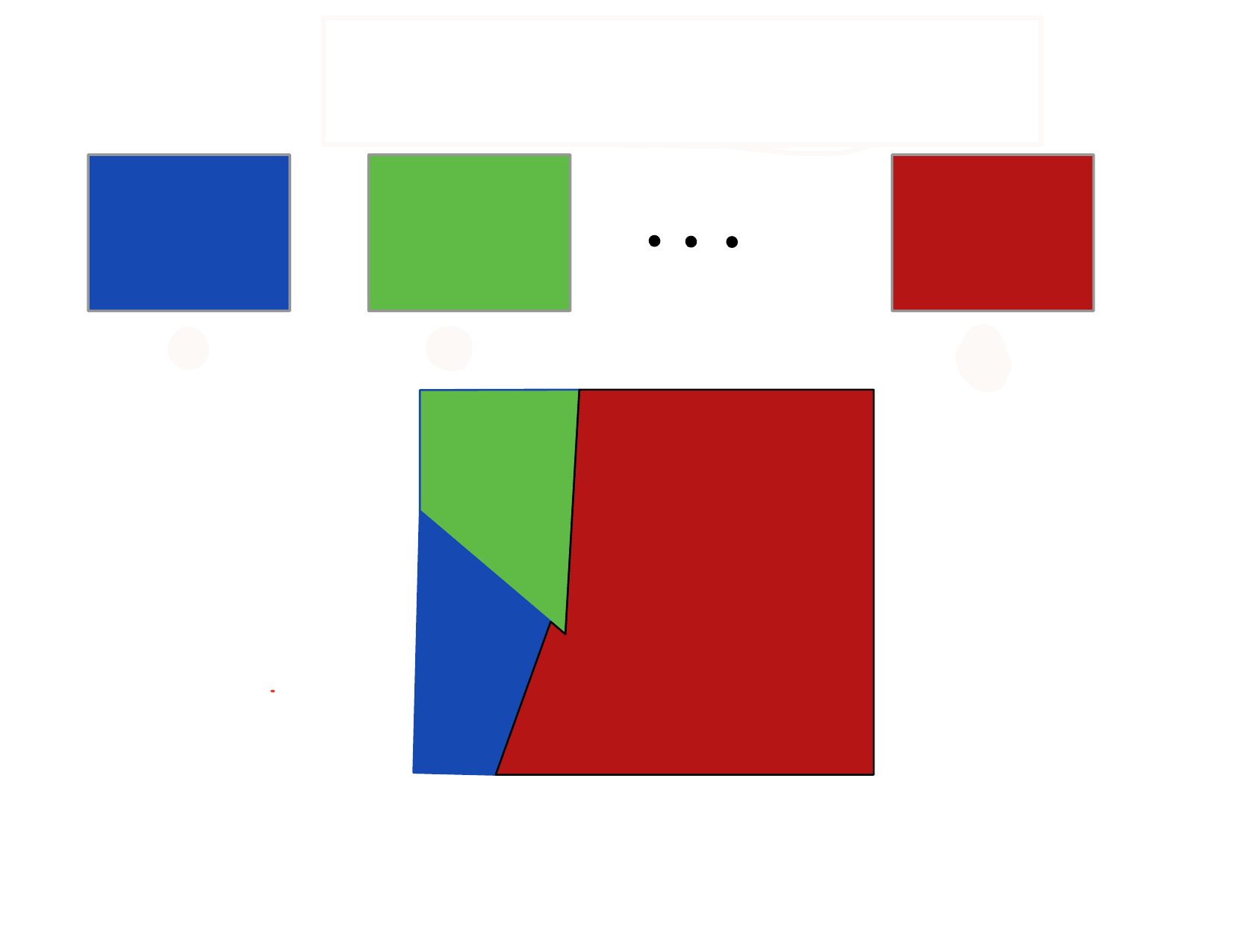}
    \caption{An ordered Potts model, its states and a mosaic of them.} \label{Potts} \end{center}
\end{figure}

There is the  temptation  to think of mosaics as follows. Consider a $p$-state Potts model, and let us call
its $p$ possible ordered phases red, blue, green,  etc. A mosaic then looks as figure 
\ref{Potts}. There is a large correlation inside each region, but their boundaries are disordered in the sense that  the fracture lines are randomly placed.

We may construct a `library' with exactly $p$ states and thus classify each `tile' of the mosaic according to its color. Here we need to pause and think: what we are storing in the library is 
an infinite - e.g. all-green -- configuration, and we are recognizing a tile by it being a fragment
of these infinite - idealized - states.

\subsection{A better one.  } \label{ff}

The pitfalls of the previous picture become evident as soon as we consider any form of non-periodic order. The example of a quasicrystal \cite{shechtman1984metallic,levine1984quasicrystals} is enough for our purposes here.

\begin{figure} 
    \begin{center}
        \includegraphics[angle=0,width=8cm]{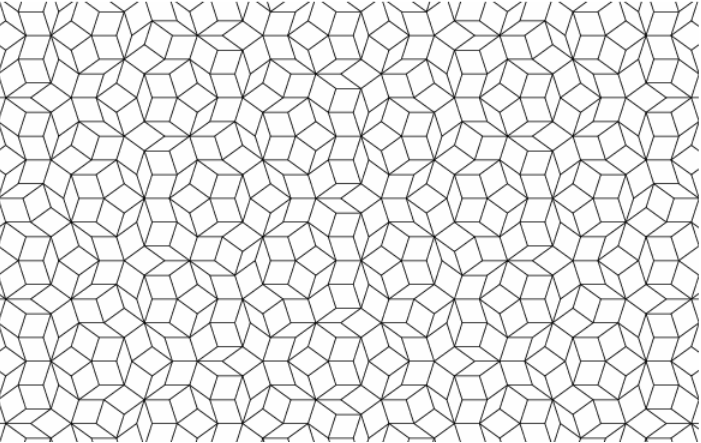}
    \caption{A Penrose tiling.}\label{penrose00}
    \end{center}
\end{figure}

\begin{figure}
    \begin{center}     \includegraphics[angle=90,width=9cm]{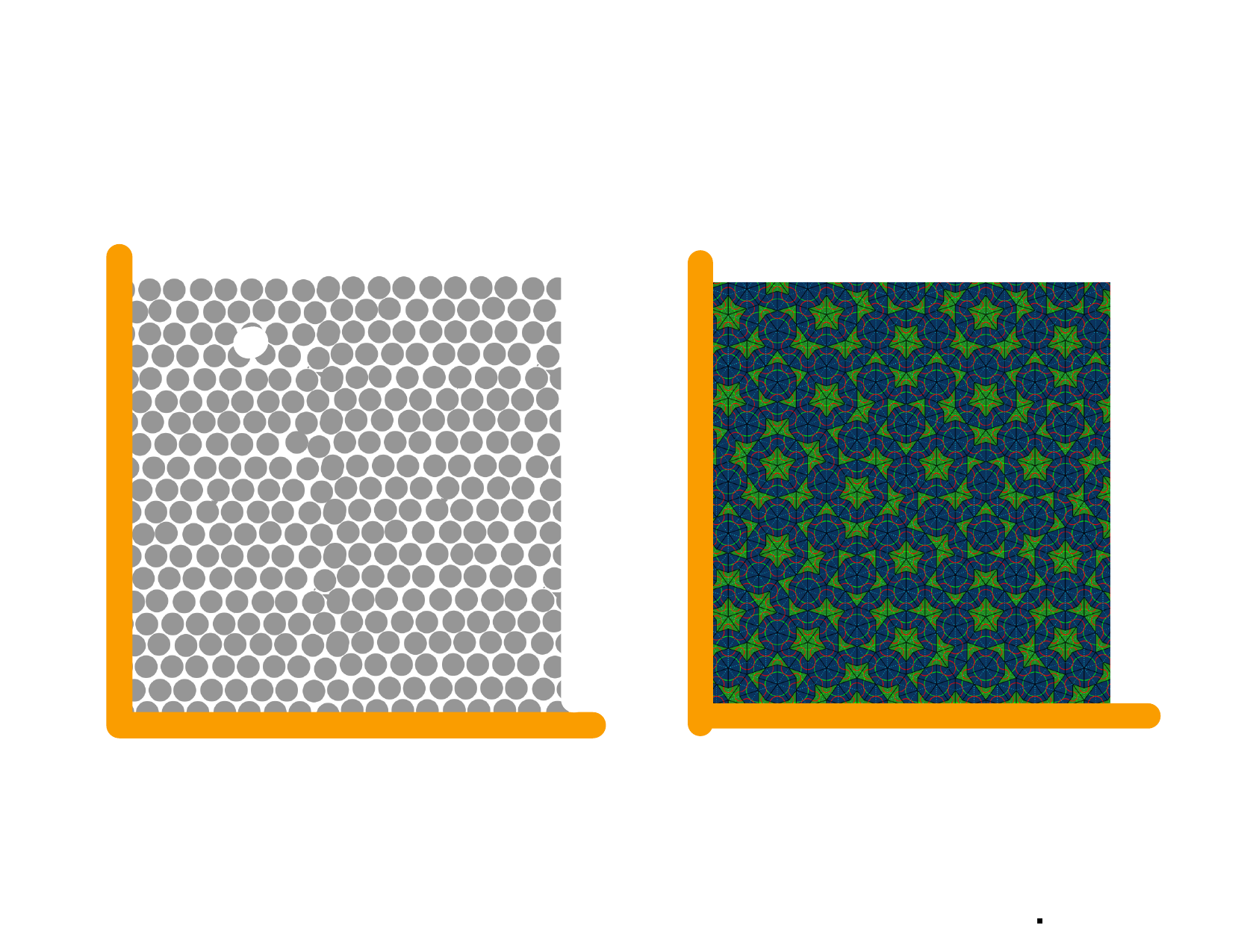}
    \caption{Boundary determines interior in a crystal and a quasicrystal} 
    \label{boundary}
    \end{center}
\end{figure}

A quasicrystal is a nonperiodic arrangement; we give here the example of a Penrose tiling \cite{grunbaum1987tilings}- see Figs  \ref{penrose00} and \ref{boundary}. We may make one such structure infinitely large, without incurring  empty spaces, i.e. `defects'. 
Consistent with the `point to set' idea, appropriate boundaries uniquely determine the interior.

\begin{figure}
    \begin{center}    \includegraphics[angle=90,width=9cm]{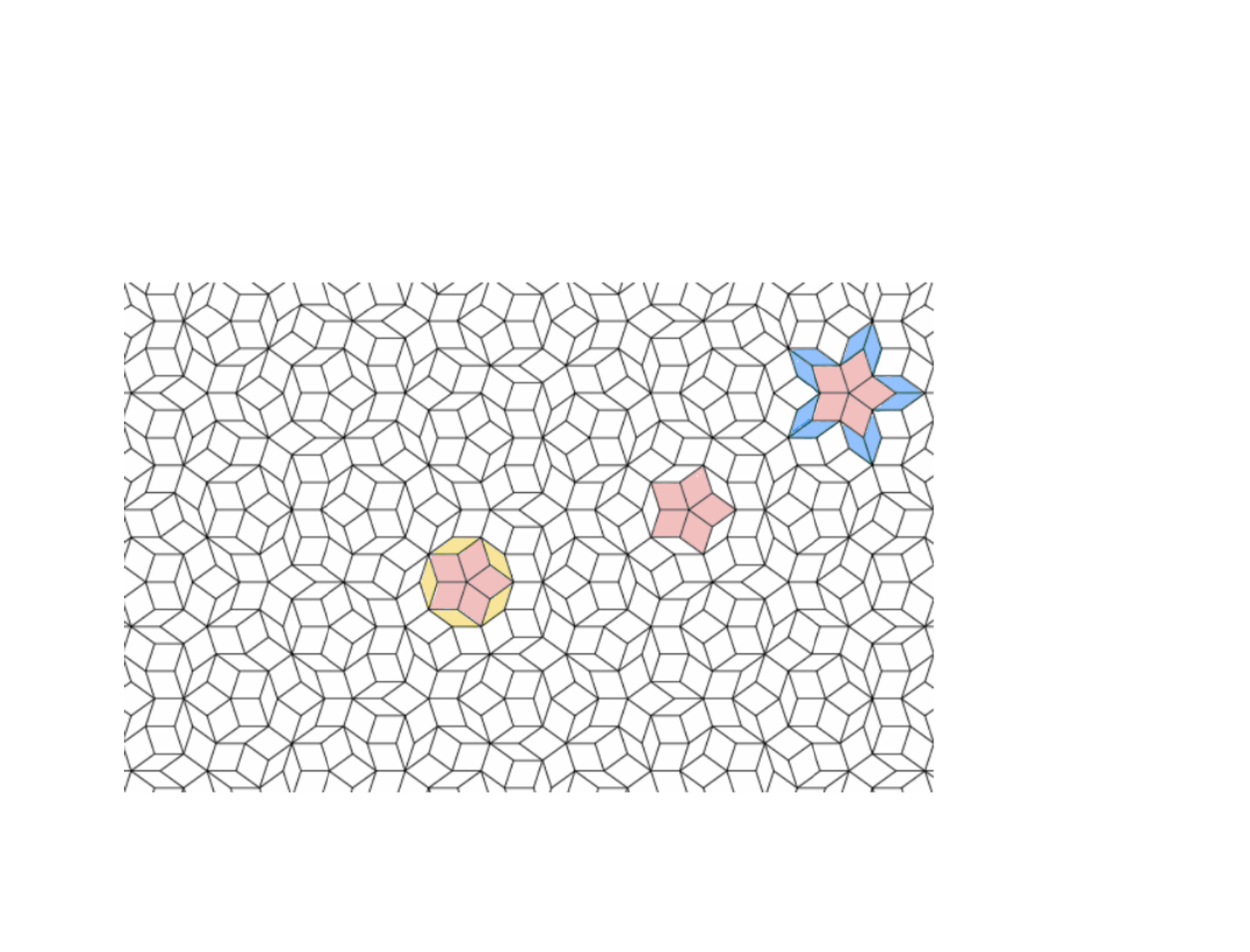}
    \caption{Frequently recurring motifs.} \label{star}\end{center} 
\end{figure}

It is well known that in a large quasiperiodic tiling, patches of every size repeat often \cite{grunbaum1987tilings} - see Fig  \ref{star}, although in a manner that does not make the system periodic. This rather surprising property is, in fact, a very
general feature of any nonperiodic arrangement: energetically `convenient' patches reappear often.

\begin{figure}
    \begin{center}       \includegraphics[angle=90,width=8cm]{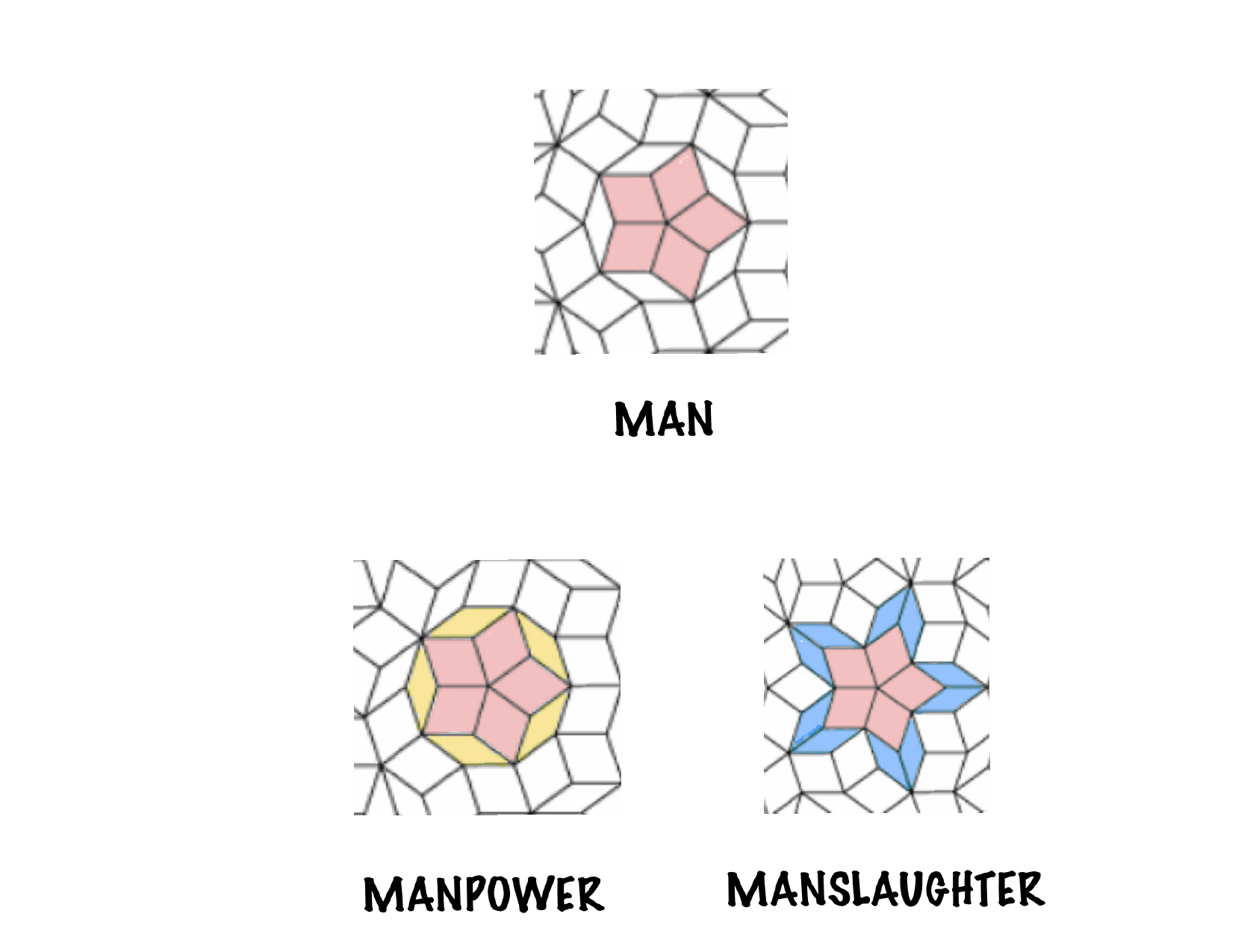}
    \caption{The `star' may be continued into a `gem' or into a `big star', just as composing words.} \label{composite} \end{center}
\end{figure}

\begin{figure}
\begin{center}
  \includegraphics[angle=90,width=10cm]{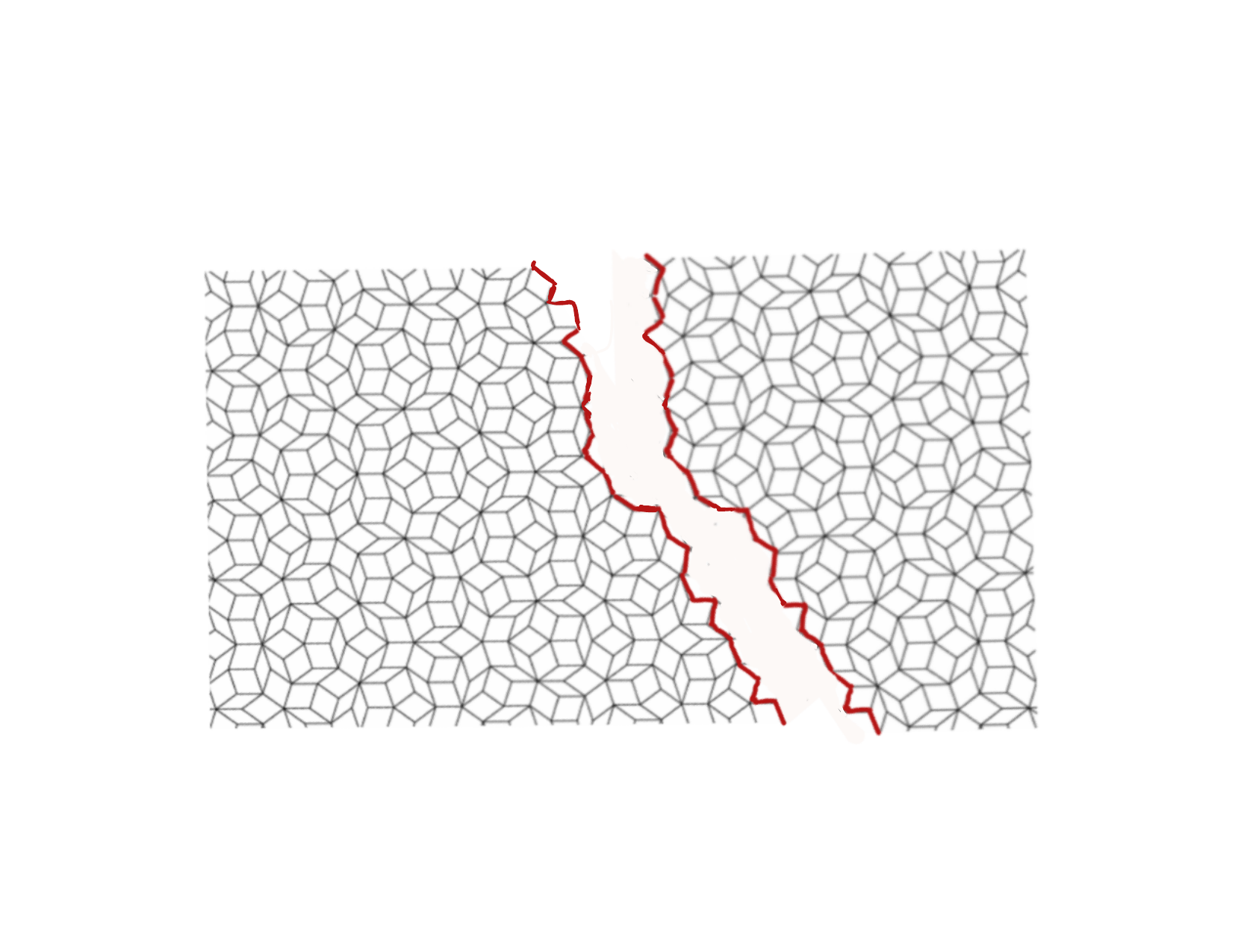}
    \caption{Two mismatched fragments of a Penrose tiling.} \label{mismatched}
    \end{center}
\end{figure}

Now consider a two-tile system as in Fig \ref{mismatched}, obtained by sticking together incorrectly two random pieces. We immediately see that our image of a `library' in the naive version of the previous section
breaks down: different fragments have many elements in common. 
What are the two `different colors'?  What shall we store in our library?  

A possible answer could be to spot the lines of defects. The problem is that defects are not
obviously lines   (see Fig 12), and furthermore, a patch near a defect may seem superficially `right', but it inevitably leads to a defect: these are called `deceptions' in the context of quasiperiodic tilings. Are deceptions ordered in spite of  not belonging to any perfect quasicrystal? Are they then a defect, even if the obvious defect is not within them, and the energetic cost is not yet paid?

\begin{figure}
\begin{center}
  \includegraphics[angle=90,width=10cm]{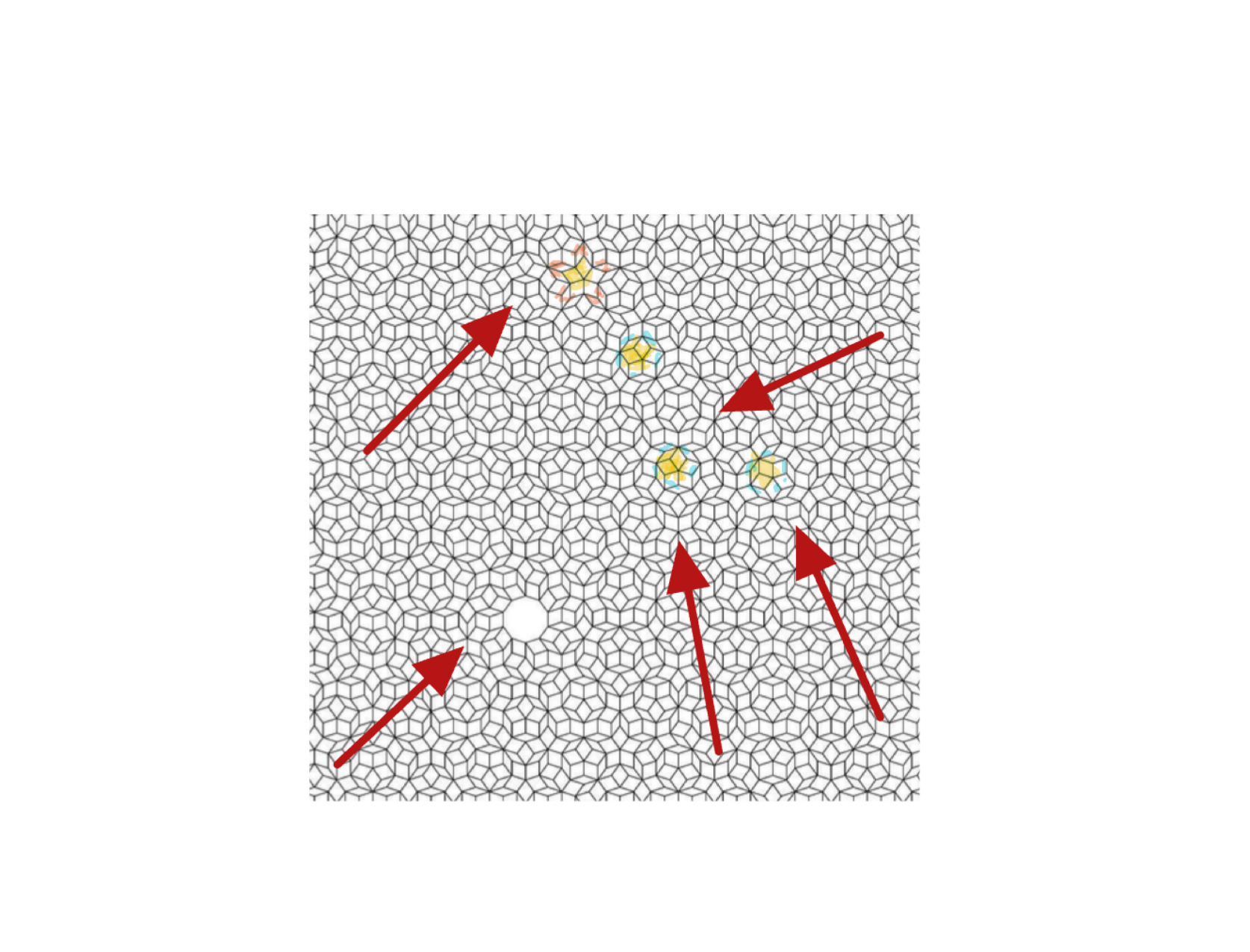}
    \caption{An  defect (low, left) that can be only cured with large rearrangements. The surroundings are what is called a `deception', despite their apparent lack of disorder, they don't belong to an infinite perfect quasicrystal.} \end{center}  \label{defect}
\end{figure}

\subsection{The structure of the library}

What size are the patches in our library? Infinite, as in the case of the Potts model above? Or finite, for example obtained by the point-to-set procedure. Both possibilities seem problematic for  constructing mosaics. 

Looking at Figure \ref{composite}, we begin to see a way out. The crucial idea is to give up attempting 
to store infinite configurations in your library and  to start constructing it bottom-up and recursively \footnote{One can insist on discussing full amorphous order in the large size limit, this requires doing it with care, introducing the concept of `metastate'. See below}.
In Fig  \ref{composite} we see how a star shape is often repeated, although sometimes it is continued into
a `gem' and sometime into a `spiky' figure. The situation is analogous to a dictionary where
composite words are allowed: from `man' we can go to `manpower' or `manslaughter'. Our dictionary should label $man=1$ , $slaughter=2$, $power=3$, $full=4$ and then $manslaugher\equiv 5=1+2$, $manpower \equiv 6=1+3$, $powerful=7=3+4$. If this article were in German, we could go to the next steps and  make composites of composites.

In this way, our library' contains building blocks, building blocks composed of smaller building blocks, and so on. If we ever hit against a defect, because of its random nature, it will be seldom repeated in the same form; the analogy would be a word with a spelling mistake,  necessarily a rare occurrence. This will be more so the larger the number of defects, as with two spelling mistakes in the same  word. 

There is nothing new in our construction: we are just following the steps to devise a compression program. Many theorems  have been proven about the optimality of such a procedure \cite{lempel1986compression,ziv1977universal,ziv1978compression,ziv1993measure}, i.e. that our dictionary has all it needs and nothing in excess.
For example, for a perfect quasicrystal, we know that the amount of information is polynomial in the number of tiles, and hence the information content (its logarithm) is subextensive in the large-size limit. Clearly, breaking this quasicrystal randomly into many pieces, there are exponentially many ways of reassembling them, and this can only be coded with an extensive information, scaling with the
exponential of the number of pieces.

What we have described here is perfectly applicable to liquids and glasses, but only once we deal correctly with  fast relaxations such as vibrations and small defects.

Before closing this section, let us remark that there is an extensive literature on the classification of the relevant mosaics for melts of particles interacting via a spherically symmetric  potential, see for example the `Topological Cluster Classification'  of Reference \cite{malins2013identification}.

\section{Treating fast degrees of freedom correctly} \label{temp}

Let us now face the problem of `factoring out' fast events, including vibrations. 

\subsection{Total entropy}

\begin{figure}
    \begin{center}
        \includegraphics[angle=90,width=8cm]{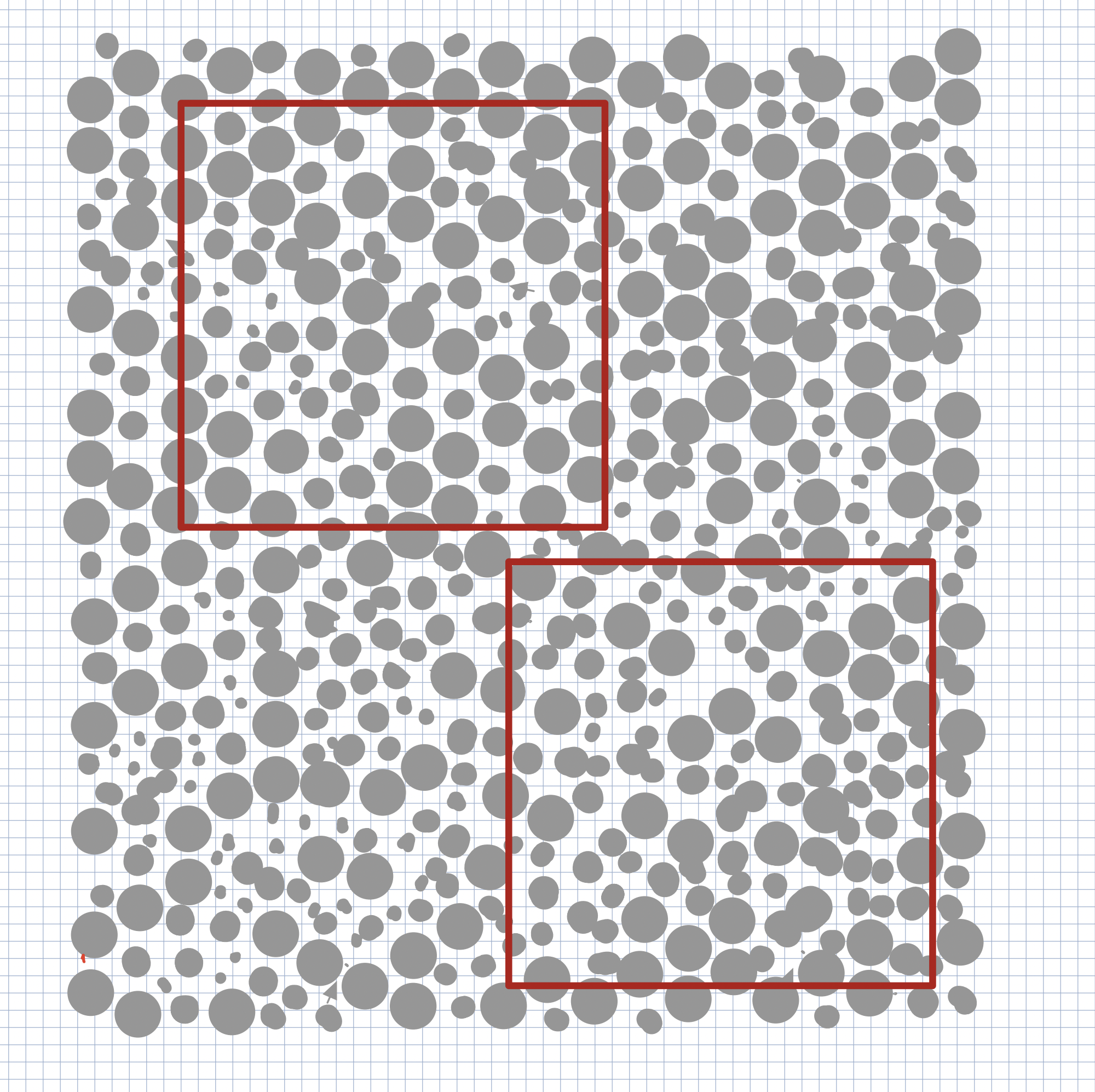}
  \caption{On a single  large system, or on many realizations of a small one   -- we compute the number of configurations, equality being defined as up to a grid of precision $\epsilon$.}  \label{entropy} \end{center}
\end{figure}

We start from the standard construction for the entropy. We consider a very large sample, and select many patches of size $L$. We classify the configurations according to a precision $\epsilon$,
as in Fig. \ref{entropy}. Our aim is to compute $S$, or, more generally, the Renyi entropies (for a detailed perspective, see Ref \cite{ozawa2024perspective}).
We consider a large sample containing $M$ different patches ${\bf a_1, ..., a_M}$ each of volume $V$. 
 Denoting by $n_a^{(V)}(\epsilon) $ the number of patches that coincide in this way with the patch ${\bf a}$, we have:
 \begin{equation}
K_q \sim -  \lim_{\epsilon \rightarrow 0} \lim_{V \rightarrow \infty} \frac{1}{ (q-1) V    } \ln \left\{ \frac{1}{M} \sum_a [n_a^{(V)}(\epsilon)]^{(q-1)} \right\}
\label{GP3}
\end{equation}
and in particular $S= \lim_{q \rightarrow 1} K_q$ is the usual entropy.
As usual with the definition of entropy, the coarse-graining parameter $\epsilon$ affects the number of patches that do coincide -- and is zero when $\epsilon \rightarrow 0$ -- but one is saved by the fact that this happens only in a subdominant way with respect to the volume $S \sim V s + a \ln \epsilon $ so the appropriate limit is well-defined. A practical computational way to extract this limit was devised by Grassberger and Procaccia \cite{grassberger1983measuring}, and is extensively used.

\subsection{Configurational entropy}

\begin{figure}
    \begin{center}
        \includegraphics[angle=90,width=8cm]{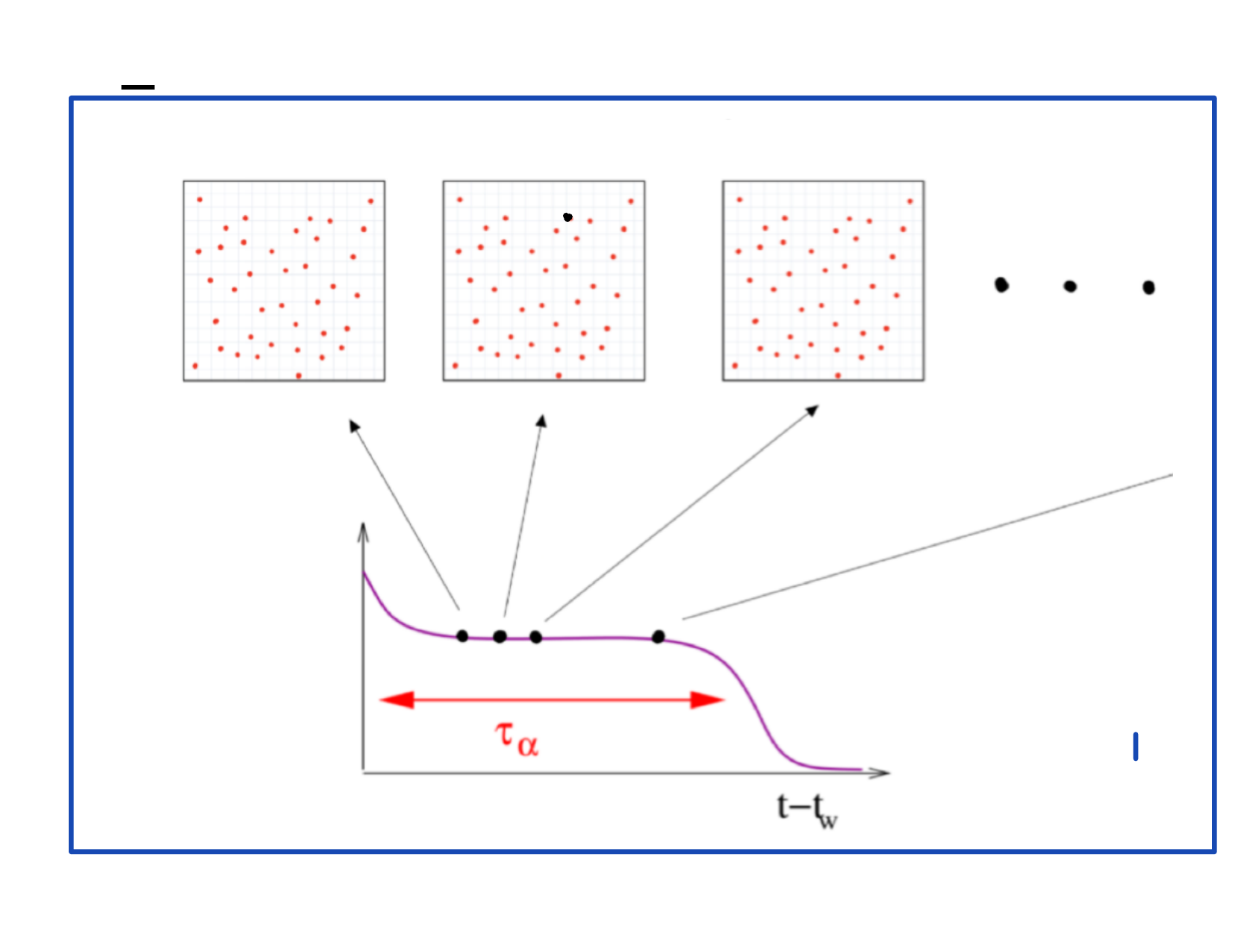}
    \caption{Computing the `vibrational' and `structural' entropies only relies on timescale-separation.} \label{14}\end{center}
\end{figure}

Solids in general, supercooled liquids, and glasses exhibit timescale-separation: there is a fast dynamics  involving vibrations or nearby particle swaps, and then there are the rearrangements whose timescale scales with the viscosity. For a crystal, this looks as in Fig. \ref{vibrations}. 
The correlation functions, on the other hand,  clearly show an extensive plateau that separate fast from `$\alpha$' scales,  as shown  in Figure \ref{14}.

The configurational entropy is a measure of entropy of all configurations, but treats as identical  all the class that are a fast rearrangement away from  each other. To compute and define  complexity properly, a strategy akin  as the one used to detect common authorship between books \cite{benedetto2002language} can be  employed:  pick  configurations at $N$  different times within the plateau
(Fig \ref{14}) and concatenate them into a a single figure.  Then,  compute the system's entropy $S_N$ obtained this way. Comparing 
the entropies of concatenated versus non-concatenated configurations we have:
\begin{equation}
    S_N = N S_{fast} + \Sigma  \hspace{1cm} ; \hspace{1cm} S= S_{fast}+\Sigma
\end{equation}
which allows us to define the complexity $\Sigma$ as the mutual entropy
\begin{equation}
 \Sigma  = \frac{ N S - S_N }{  (N-1)} 
\end{equation}

The same definition may be used for the Renyi entropies. Also very important: {\it we have never assumed equilibrium here}, so the definition holds even for an aging glass.

\begin{figure}
    \begin{center}
       \includegraphics[angle=0,width=9cm]{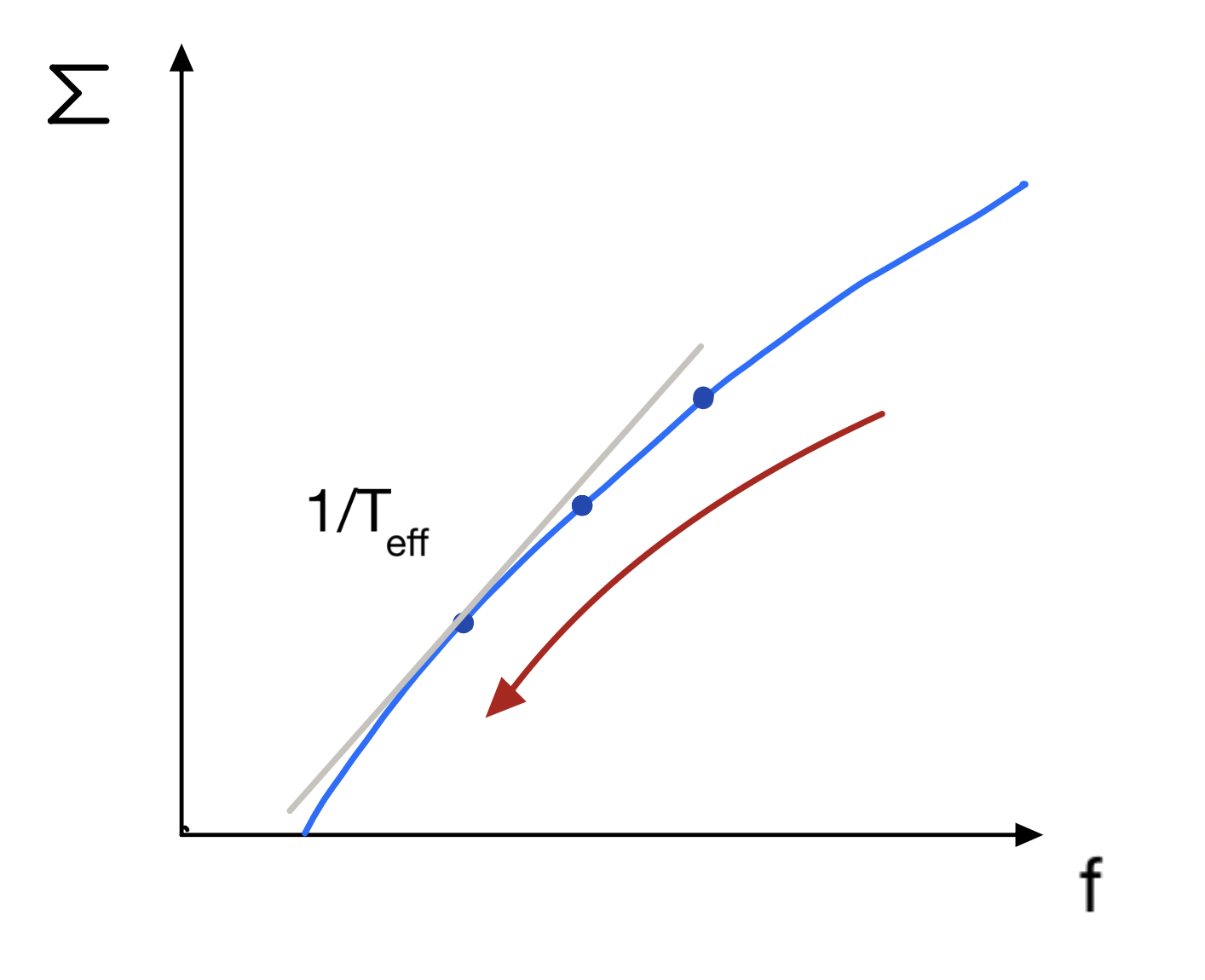}
       \vspace{0cm}
    \caption{Effective temperature or replica parameter $X$ in an aging situation } \label{config}\end{center}
\end{figure}

\section{Physical situations and definitions \label{definitions}} 

The mathematical literature discusses  the definition of pure states in equilibrium and sending the system size  to infinity, a construction that need be done with care. The procedure is the `metastate construction' \cite{newman2022metastates,read2022complexity}. 

From a pragmatic standpoint, experiments involve quenching systems  from high temperatures to a  low (possible zero) temperature bath. At all finite times the boundary conditions are
irrelevant, provided they are sufficiently (i.e. a few correlation lengths)  far away, a condition that can  be checked at each step -- the statistics are ideally obtained by repeating the experiment.  
  The complexity, and from it the typical length,  may be evaluated also
 in this out of equilibrium situation. We will see the former decrease slowly, and possibly saturate at some point in time. The question of the existence of an ideal glass thus becomes straightforward: would this length continue to grow indefinitely with more time,  or does it always eventually reach a  saturation point? The reason for the stop of the growth could stem from  the system equilibrating  at  finite time, or the nucleation of a crystal making the glass  irrelevant, or yet that another form of transition intervenes. 

We have no answer to this question. However, from the physical reality point of view one must realize that the lengths involved in real glasses are exceedingly short; only a few molecules are correlated. 
The estimates  made in the point-to-set framework are done in polydisperse systems by refilling the cavity with particles of the same sizes as the original ones, rather than randomly sampling them from the distribution. This is legitimate, but it clearly leads to longer PTS lengths. On the other hand, freezing an out of equilibrium configuration,  the PTS length may be dramatically longer than the entropy-related one for the same situation. One can easily understand this in the case of hard spheres crunched to high density or in general at very low temperatures: introducing the same particles,  the chances that another variant of the configuration contributing to equilibrium inside the cavity are vanishing.

\section{The growth of correlations, Renyi complexities and Effective Temperatures \label{renyi}} 

A picture that has emerged of a glass, based on the mean field solution \cite{cugliandolo1993analytical,cugliandolo1997energy} but also with important precedents at the phenomenological level \cite{tool1946viscosity}, is as follows. The system is taken at high temperature and put in contact with a low temperature bath.  The landscape is composed of a large number of metastable states, distributed in terms of their internal free energy as in Figure \ref{renyi}. As the liquid relaxes, it does so exploring the states at a given level {\it democratically}: this remarkable fact can be rationalized by realizing that every new relaxation is very much slower than the preceding ones. The probability distribution of a state `$a$' is then approximated by
\begin{equation}
p_a \propto e^{-\frac{1}{T_{eff}} f_a}\label{monasson}
\end{equation}
where $T_{eff}$ slowly decreases with time, until finally reaching $T_{eff}=T$. This is the construction of Monasson \cite{monasson1995structural}.
Because fast relaxations occur at the bath temperature, but slow ones are essentially thermalized at a slowly decreasing
effective temperature, this explains why,  at  low frequencies the fluctuations and dissipations are related \cite{cugliandolo1997energy}
by $T_{eff}$ precisely  (see the construction in Fig \ref{renyi}). 
Clearly, this is only literally true for mean field systems, but there is abundant evidence that the scenario holds, at least qualitatively, in realistic glasses.

Now, comparing equation (\ref{monasson})
with the definition of Renyi entropies (\ref{GP3}) -- the latter being applicable to realistic glasses -- we see that the  configurations that dominate  the Renyi entropies $K_q$ are those that dominate also at times such that:
\begin{equation}
    T_{eff} = \frac{T}{q}
\end{equation}
This establishes a connection between the configurations that dominate the  Renyi entropies calculated in equilibrium, and those explored by the dynamics on their way to equilibrium. `Higher' states dominate for small $q$, but these are those that have higher configurational entropies, and hence smaller mosaics.

\section{Conclusion}

In conclusion, thinking in terms of information and information storing 
is not an elegant curiosity, but rather a necessity for the description  liquids and glasses.

\pagebreak

\bibliographystyle{unsrt}

\bibliography{toulouse}

\end{document}